\documentclass[sigconf]{acmart}

\copyrightyear{2023}
\acmYear{2023}
\setcopyright{acmlicensed}\acmConference[ICTIR '23]{Proceedings of the 2023 ACM SIGIR International Conference on the Theory of Information Retrieval}{July 23, 2023}{Taipei, Taiwan}
\acmBooktitle{Proceedings of the 2023 ACM SIGIR International Conference on the Theory of Information Retrieval (ICTIR '23), July 23, 2023, Taipei, Taiwan}
\acmPrice{15.00}
\acmDOI{10.1145/3578337.3605137}
\acmISBN{979-8-4007-0073-6/23/07}

\makeatother

\usepackage{enumitem}
\newlist{inlinelist}{enumerate*}{1}
\setlist*[inlinelist,1]{%
  label=(\roman*),
}
\usepackage{subfigure}
\usepackage{multirow}
\usepackage{balance}
\usepackage{tablefootnote}

\usepackage{xspace}

\title{Pre-Training Multi-Modal Dense Retrievers for Outside-Knowledge Visual Question Answering}

\author{Alireza Salemi}
\affiliation{\institution{University of Massachusetts Amherst}
\country{United States}}
\email{asalemi@cs.umass.edu}

\author{Mahta Rafiee}
\affiliation{\institution{University of Massachusetts Amherst}
\country{United States}}
\email{mrafiee@cs.umass.edu}

\author{Hamed Zamani}
\affiliation{\institution{University of Massachusetts Amherst}
\country{United States}}
\email{zamani@cs.umass.edu}

\begin{document}


\begin{abstract}
This paper studies a category of visual question answering tasks, in which accessing external knowledge is necessary for answering the questions. This category is called outside-knowledge visual question answering (OK-VQA). A major step in developing OK-VQA systems is to retrieve relevant documents for the given multi-modal query. Current state-of-the-art asymmetric dense retrieval model for this task uses an architecture with a multi-modal query encoder and a uni-modal document encoder. Such an architecture requires a large amount of training data for effective performance. We propose an automatic data generation pipeline for pre-training passage retrieval models for OK-VQA tasks. The proposed approach leads to 26.9\% Precision@5 improvements compared to the current state-of-the-art asymmetric architecture. Additionally, the proposed pre-training approach exhibits a good ability in zero-shot retrieval scenarios.
\end{abstract}






\maketitle

\section{Introduction}

\begin{figure}[t]
    \centering
    \includegraphics[width=\linewidth]{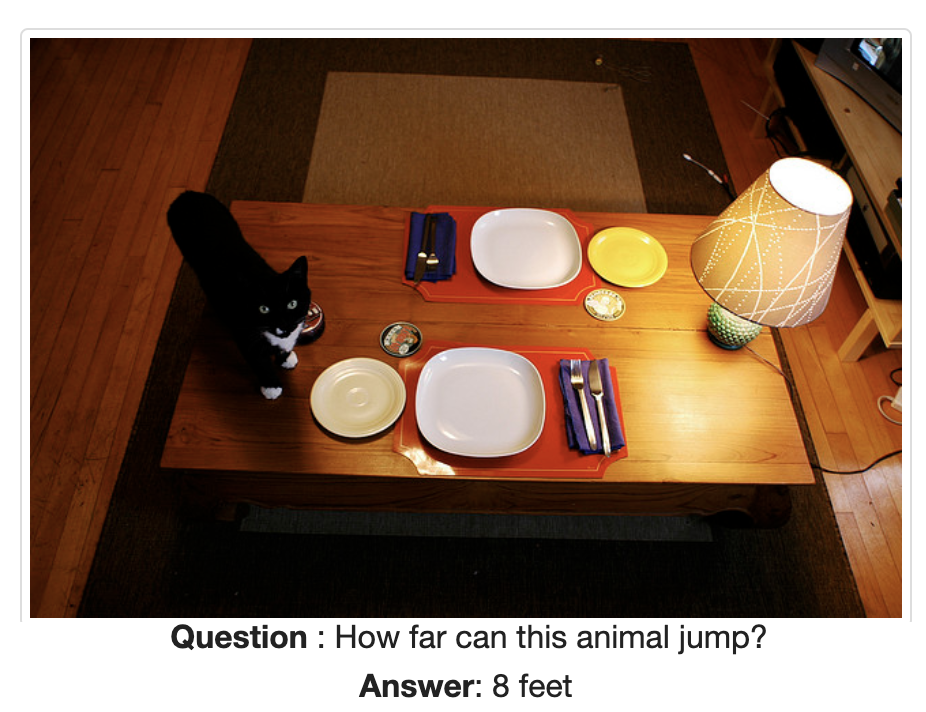}
    \caption{An example OK-VQA question. Answering this question requires external knowledge. \\     {\normalfont\footnotesize Image \copyright~ nsfmc,
        \url{https://www.flickr.com/photos/subliminal/589841807}}} 
    \label{fig:example}
\end{figure}

Outside-knowledge visual question answering (OK-VQA) \cite{okvqa} is a category of visual question answering tasks in which answering the given natural language question about an image requires access to external information. In OK-VQA retrieval tasks, queries are multi-modal (text and image) and the retrieval corpus is often uni-modal, consisting of text documents. To shed more light on this, Figure \ref{fig:example} presents an example of the queries in this task. It can be observed that answering the question `how far can this animal jump?' requires an understanding of the entity (i.e., the cat) in the image, but this information may not be sufficient. In this case, accessing a knowledge source containing information about the animal in the picture and its abilities can facilitate answering the question. The extensive range of practical implementations for OK-VQA is worth noting. Consider the scenarios where individuals, who are patrons of e-commerce platforms, capture images of products or specific components and pose queries about them \citet{salemi2023symmetric}. Similarly, within the educational domain, students can interrogate an image from their textbook by asking questions \citet{salemi2023symmetric}. Moreover, users can leverage OK-VQA by photographing visual signs or artwork and inquiring about their significance or historical background. These instances merely scratch the surface of the diverse application potential of OK-VQA.

Lately, \citet{salemi2023symmetric} introduced a symmetric architecture for multi-modal retrieval and compared it with the previous state-of-the-art asymmetric architectures introduced by \citet{okvqa-passage}. Despite the significantly better performance of the symmetric architecture, the fact that this architecture needs access to a caption generator at the inference time makes it costly to use in real-time. Consequently, the main emphasis of this research paper is placed on asymmetric architecture, unveiling a novel methodology for enhancing the training of superior asymmetric retrievers. Importantly, this approach effectively curtails the necessity of caption generation solely to the training phase, sparing it from being required during inference time.

\citet{okvqa-passage} demonstrates that supervised asymmetric dense retrieval models with multi-modal query encoder and uni-modal document encoder lead to state-of-the-art passage retrieval performance for OK-VQA tasks. However, it requires large-scale manually labeled training data which is expensive and time consuming to obtain. Inspired by the prior research on text retrieval based on weak supervision \cite{Dehghani:2017:weak,WeakTheory,RELWE} and Inverse Cloze Task (ICT) pre-training \cite{ict}, this paper introduces a novel pipeline for automatic generation of training data for OK-VQA tasks. This data generation pipeline requires no manually labeled OK-VQA data. It first obtains an image corpus (e.g., MS COCO \cite{mscoco}) and generates captions for the images. Each caption is then used as a query to retrieve text passages from Wikipedia. We then select some noun phrases from each passage as potential answers and generate a question for each of them using a fine-tuned language model. To reduce the noise introduced into the pre-training data, we design a question-answering model and filter out questions for which the model cannot produce a close enough answer. This process leads to a large-scale dataset with about 4.6 million question-image pairs for OK-VQA tasks. The generated data can then be used for pre-training dense retrieval models for OK-VQA tasks. To the best of out knowledge, this is the first attempt to automatic generation of data for OK-VQA tasks.

Our experiments on the OK-VQA passage retrieval dataset \cite{okvqa,okvqa-passage} demonstrate that training dense retrieval models using the proposed data generation pipeline leads to 40.2\% Precision@5 improvements in a zero-shot setting compared to competitive baselines. We also show that pre-training state-of-the-art supervised dense retrieval models improves state-of-the-art performance by 26.9\% in terms of Precision@5. The obtained improvements are statistically significant in all cases. Further analysis suggests that the proposed pre-trained model that is fine-tuned only on 25\% of the OK-VQA supervised data outperforms the model that is trained on 100\% of the supervised data without pre-training. Moreover, the performance of the pre-trained model becomes relatively stable after observing 50\% of the supervised training data. Therefore, the proposed pre-training procedure reduces the need to large-scale manually labeled training sets. 

In summary, the major contributions of this work include:
\begin{enumerate}[leftmargin=*]
    \item Introducing the first automatic data generation pipeline for outside-knowledge visual question answering tasks.
    \item Improving the current state-of-the-art asymmetric passage retrieval models in both zero-shot and supervised settings.
    \item Providing extensive result analysis to better understand the impact of pre-training on OK-VQA performance. 
\end{enumerate}

To foster research in this area, we release our generated dataset, our data creation pipeline, and our learned model parameters.\footnote{The data and code are available at \url{https://github.com/alirezasalemi7/pretraining-multimodal-dense-retriever-for-okvqa}}

\section{Related Work}
\label{sec:related-work}

\subsubsection*{\textbf{Multi-Modal Dense Passage Retrieval}}

Multi-modal dense retrieval can be defined in different categories based on where the multi-modality takes place. The multi-modality can be in the queries, with a corpus of uni-modal documents, which enables the underlying information need to be expressed through a multi-modal representation \cite{okvqa}. Our work fits into this category with queries comprised of images with corresponding questions, and uni-modal textual passages in the corpus. Another line of work has been focusing on multi-modal documents in the corpus, such as a mix of textual, tabular, or visual information, while the query is expressed in one modality \cite{manymodal, talmor2021multimodalqa, https://doi.org/10.48550/arxiv.2209.00179}. In another setting, both queries and documents can be multi-modal, for example where the answer to a query about an image contains multiple modalities \cite{singh-etal-2021-mimoqa}. Cross-modal retrieval is also partly related to multi-modal retrieval, where both queries and documents are uni-modal but they come from different modalities \cite{align, clip}.

\subsubsection*{\textbf{Outside-Knowledge Visual Question Answering}}

In standard visual question answering (VQA) \cite{vqa}, the answer lies in the image; however, in outside-knowledge visual question answering (OK-VQA) \cite{okvqa}, the image and question are jointly used to find the answer to the question from an external knowledge source \cite{okvqa-passage}. That being said, retrieving relevant passages to a query, which consists of an image and a question about it, plays an essential role in this task \cite{okvqa-passage}. Previous work \cite{conceptbert, unifer, rvl, krisp,mavex} mostly utilizes knowledge graphs as a source of external information; however, the lack of a complete and easily updatable knowledge source is challenging \cite{ABUSALIH2021103076, tang-etal-2019-learning}. Therefore, following \citet{okvqa-passage} and \citet{salemi2023symmetric}, we focus on retrieving passages from Wikipedia as the knowledge source. 

Previous work mostly evaluates OK-VQA based on the answer generation quality \cite{conceptbert, unifer, rvl, krisp,mavex, kat, pica, lako}; however, following \citet{okvqa-passage}, we only investigate the retrieval performance in the aforementioned task. In contrast with \citet{salemi2023symmetric}, which focuses on designing a symmetric architecture for OK-VQA retrieval and answer generation, we investigate the data generation and augmentation methods to train the proposed asymmetric retriever architecture by \citet{okvqa-passage} with no labeled training data.

\subsubsection*{\textbf{Pre-Training Dense Passage Retrievers}}

In recent years, pre-training transformers \cite{transformer} using semi- and self-supervised tasks has become a standard approach for achieving strong performance in natural language and vision tasks \cite{bert, roberta, vit}. Moreover, retrieval-specific pre-training tasks, such as Inverse Cloze Task (ICT) \cite{ict}, have been shown to be effective for uni-modal retrieval. Recently, a multi-modal variant of ICT has been proposed by \citet{Multi_ICT}, in which queries are question-image pairs, and documents are passage-image pairs. However, our work focuses on the case that passages are only textual, while queries consist of question-image pairs. 

The research by \citet{pre-train-vqa} is perhaps the closest work to ours, in which the authors focus on pre-training models for VQA tasks, which is by nature different from OK-VQA. \citet{pre-train-vqa} only generates questions from the image captions due to the nature of VQA, in which the answers lie in the image. In contrast, we use captions to retrieve a relevant passage to the image and generate questions from that passage to ensure that answering them requires external knowledge.

\section{Problem Statement}

While multi-modal retrieval can be defined in different ways as mentioned in section \ref{sec:related-work}, this paper only focuses on multi-modal scenarios where the query $(Q, I)$ consists of the question $Q$ about the image $I$, and the corpus $C$ from which relevant passages should be selected is only textual.

Suppose $T = \{(Q_1, I_1, A_1, R_1), ..., (Q_N, I_N, A_N, R_N)\}$ represents the training set for multi-modal retrieval in this paper. Each training sample in $T$ consists of a question $Q_i$ written in natural language, an image $I_i$, a set of answers $A_i$ to the question $Q_i$, and a set of relevant passages $R_i$ that contains the answer to $Q_i$. In more detail, the answer set $A_i$ might contain more than one answer to the question, which are syntactically different but semantically the same ($|A_i| \geq 1$). Additionally, each question and image might have more than one related passage ($|R_i| \geq 1$ and $R_i \subseteq C$).

The main task in this paper is to use training set $T$ to train a dense retriever that takes query $(Q, I)$ as input and retrieves $K$ passages that are relevant to the query from the corpus $C$ ($|C| \gg K$). In this paper, we introduce a pipeline for generating weakly supervised data, similar to the proposed problem definition, to first pre-train the model on the weakly supervised generated data and then fine-tune the pre-trained on the task's data. The following sections explain our proposed pipeline for this purpose.

\section{The Proposed Pre-training Pipeline}
Automatic data generation for (pre-)training neural models for text retrieval and question answering has proven to be effective. For instance, \citet{Dehghani:2017:weak} introduced weak supervision in information retrieval by utilizing an existing unsupervised retrieval model as a weak labeler. \citet{WeakTheory} provided theoretical justification on when and why weak supervision lead to strong and robust improvements. \citet{GPL} used a similar approach for adapting well-trained retrieval models to an unseen target domain. More recently, \citet{ict} used Inverse Cloze Task for pre-training text retrieval models and \citet{InPars} used large-scale language models, such as GPT \cite{Radford2019LanguageMA}, for data generation. All these approaches are developed for text retrieval tasks.

For multi-modal tasks, \citet{pre-train-vqa} focused on pre-training models for visual question answering (VQA) tasks, which is fundamentally different from OK-VQA. \citet{pre-train-vqa} only generates questions from the image's caption due to the nature of VQA, in which the answers lie in the image (e.g., asking about the color of an object in the image). VQA is not an information-seeking task; thus, this approach cannot be applied to OK-VQA.

This section introduces our data generation pipeline for pre-training dense passage retrieval models for OK-VQA tasks. 

\begin{figure*}
    \centering
    \includegraphics[width=\linewidth]{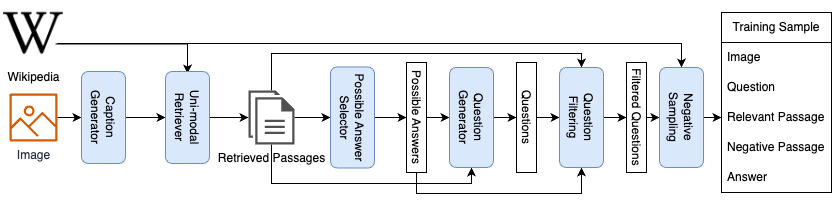}
    \caption{The proposed data generation pipeline for pre-training OK-VQA models.} 
    \label{fig:data-generation-pipeline}
\end{figure*}

\subsection{Automatic Data Generation for Pre-training}
Figure \ref{fig:data-generation-pipeline} depicts an overview of our automatic data generation pipeline for pre-training multi-modal dense passage retrieval models. We start with an image and use an automatic image captioning model to produce a textual description of the image. We then retrieve $M$ passages from a large collection, such as Wikipedia, given the image caption as the query. We then extract a set of potential short answers from the retrieved passages. For each potential answer, we generate a question using a sequence-to-sequence model. We later filter out low quality questions. A negative selection component is also developed to produce data for optimizing retrieval models. The outcome of our pipeline is a set of data instances, each represented as $(Q_i, I_i, A_i, R_i, N_i)$, where $Q_i$ is a question about the image $I_i$, $A_i$ is the answer to the question $Q_i$, $R_i$ is a relevant passage to the question, and $N_i$ is a hard negative passage for the question $Q_i$. In the following sections, we explain the procedure of generating each component in detail.

\subsubsection*{\textbf{Matching Images and Passages using Captions}}
For each image in the MS COCO \cite{mscoco} training set (\~82K images), we aim at retrieving $M$ passages. Therefore, designing retriever $R_{img2text}$, which takes an image as input and retrieves a set of related passages from corpus $C$), is required. We use a Wikipedia dump with 11M passages as the corpus $C$.\footnote{This Wikipedia dump is available at: \url{https://ciir.cs.umass.edu/downloads/ORConvQA/all_blocks.txt.gz}} We retrieve $M = 5$ passages for each image.

While some models, such as CLIP \cite{clip} and ALIGN \cite{align}, are designed to act as $R_{\text{img2text}}$, for simplicity and without losing generality, we use BM25 \cite{bm25} as $R_{\text{img2text}}$, in which we use a textual description of the image to retrieve a set of passages. To calculate the similarity score between the image $I$ and the passage $P$, we use the following formula: $S_{R}(I, P) = S_{BM25}(\phi_{I \rightarrow T}(I), P)$, where $\phi_{I \rightarrow T}$ is a modality converting module that takes an image and generates a textual description for it. 

Generating a description of an image can happen in several ways. For instance, the textual label of objects in an image can be used to describe the image using text. This approach suffers from two issues: 1) object labels are limited to a pre-defined set, and 2) labeling objects in images in large scale is costly. Conversely, using captions as the image description resolves the mentioned issues by generating an open-ended textual description of an image and using the large-scale available image-caption data on the web. That being said, we use ViT-GPT \cite{vit-gpt}, a transformer-based \cite{transformer} image-to-text model, to generate a caption for each image. ViT-GPT is trained on the images and captions provided by MS COCO \cite{mscoco} dataset using a cross-entropy loss function. Once the model is trained, we freeze the model's parameters and use it in inference mode. 

\subsubsection*{\textbf{Selecting Potential Answer Phrases from Retrieved Passages}}

Investigating the OK-VQA dataset shows that approximately 80\% of the answers in this dataset are noun phrases. Following this observation, we use noun phrases in the retrieved passages as potential answers. This approach has been previously used by \citet{lee-etal-2021-qace-asking}. In more detail, we use spaCy\footnote{\url{https://spacy.io/}} to extract noun phrases from passages. We consider all noun phrases as potential answers, except those that have a pronoun or determiner (e.g., "a", "an", and "the") in their subtrees. This is because pronouns and determiners usually refer to a specific word in the passage (i.e., co-references), and we would like to select ``standalone'' answer phrases. 

\subsubsection*{\textbf{Question Generation and Filtering}}
The next step in the pre-training data generation pipeline is generating a question for each selected answer phrase. Suppose $M_{QG}(A, P)$ is a question generator that takes passage $P$ and the answer phrase $A$ as input and generates a question $Q$ whose answer is $A$. To implement $M_{QG}$, following \citet{question-generation}, we feed the passage $P$ and the potential answer phrase $A$ to T5-large \cite{t5} and instruct the model to generate a question. To this aim, we utilize SQuAD v1.1 \cite{squad} dataset for fine-tuning this question generation model. For each training sample in the SQuAD v1.1 dataset, the answer is surrounded with \texttt{<hl>} token, and the passage with the surrounded answer is fed to T5. The cross-entropy loss is used for training the model:  
\begin{equation}
    L_{QG} = - \sum_{i}^{|Q|} \log{P(y_i|y_{k<i};P')}
\end{equation}
\noindent
where $y_i$ is the $i$\textsuperscript{th} token in the question $Q$, and $P'$ is the passage $P$ with \texttt{<hl>} surrounding tokens\footnote{The checkpoint for this question-generation model is available at: \url{https://huggingface.co/lmqg/t5-large-squad-qg}}.

As a reference on the quality of the question generation model, we evaluate it on the test set of SQuAD \cite{squad} and it achieves a BLEU-4 \cite{bleu} score of 27.21 and rouge-L \cite{rouge} of 54.13.
To further reduce the amount of noise in the generated pre-training data, we filter out the questions that a question-answering model cannot answer. Suppose $M_{QA}(Q, P)$ is a question-answering model, which takes the question $Q$ and the passage $P$ as inputs and generates or selects an answer phrase. Finally, we only select the generated questions that satisfy the following condition: $\textsc{Rouge-1}(A, M_{QA}(M_{QG}(A, P), P)) > T$, 
where $\textsc{Rouge-1}$ is the rouge-1 score \cite{rouge}, $A$ is the potential answer from the passage $P$, and $T$ is a threshold for the similarity of the potential answer and the answer selected by the question-answering model. We use $T = 0.5$ in our experiments.

To implement $M_{QA}$, we use a RoBERTa-base \cite{roberta} that is fine-tuned for answer span selection trained on the SQuAD dataset. The model is trained based on the log-likelihood of predicting the correct start and end tokens. For selecting the answer span, the span with the highest $P(S_i|P;Q) + P(E_j|P;Q)$ is selected where $P(S_i|P;Q)$ shows the probability of the $i$\textsuperscript{th} token being the start of the span and $P(E_j|P;Q)$ shows the probability of the $j$\textsuperscript{th} token being the end of the span. As a reference, this question-answering model achieves a F1 score of 82.91\% and exact match of 79.87\% on the test set of SQuAD v2 dataset \cite{squadv2}.\footnote{The checkpoint for this question-answering model is available at: \url{https://huggingface.co/deepset/roberta-base-squad2}}

\subsubsection*{\textbf{Negative Passage Sampling}}
Using hard negatives and their quality plays an essential role in the final performance of dense passage retrieval models \cite{dpr}. For each generated question, we retrieve passages using BM25. We choose the highest scored passage that does not contain the answer $A$ as the negative passage.

\subsubsection*{\textbf{Summary}}
The proposed pipeline leads to 4,621,973 question-image pairs from 82,783 unique images of MS COCO \cite{mscoco}. The average question, passage, and answer length in the created dataset are $9.6 \pm 3.0$, $187.2 \pm 105.7$ and $2.3 \pm 1.2$ words, respectively.


\subsection{Dense Retrieval Model}
\label{sec:arch}
The nature of the multi-modal retrieval task that we attempt to solve in this paper requires the bi-encoder dense passage retriever to encode queries in multi-modal semantic space and to encode passages in textual semantic space. We use an asymmetric state-of-the-art dense passage retrieval for OK-VQA tasks proposed by \citet{okvqa-passage}. It uses an asymmetric dense passage retriever with the multi-modal query encoder $E_{MM}$ and the textual passage encoder $E_T$. Then, the relevance score is calculated as follows: $S((Q, I), P) = E_{MM}(Q, I) \cdot E_T(P)$,
where $\cdot$ denotes the inner product. Following \citet{okvqa-passage}, we implement $E_T$ using the representation of the \texttt{[CLS]} token provided by a BERT-base \cite{bert} model. Similarly, we utilize the representation of the \texttt{[CLS]} token generated by LXMERT \cite{lxmert}, a vision-language model pre-trained with various vision-language tasks.

To train the retriever, we use a contrastive loss as follows:
\begin{equation}
    \label{eq:contrastive-loss}
    L_{DR} = - \log{\frac{e^{S((Q,I),P_{pos})}}{e^{S((Q,I),P_{pos})} + \sum_{P' \in \mathbf{P_{neg}}} e^{S((Q,I),P')}}}
\end{equation}
where $P_{pos}$ is a positive (relevant) passage and $\mathbf{P_{neg}}$ is a set of negative passages for the question-image pair $(Q,I)$. In addition to the selected negative passages, we use in-batch negatives, in which all the positive and negative passages of other queries in the same training batch are considered as negative passages to the query. We use the Faiss library \cite{faiss} for indexing and efficient dense retrieval.

\begin{table*}
    \centering
    \caption{Passage retrieval performance on the OK-VQA dataset \cite{okvqa}. The superscript $^*$ denotes statistically significant improvement compared to all baselines based on two-tailed paired t-test with Bonferroni correction ($p < 0.05$).}
    \begin{tabular}{l|cc|cc|cc|cc}
    \multirow{3}{*}{\textbf{Model}} & \multicolumn{4}{c|}{\textbf{Zero-Shot Performance}} & \multicolumn{4}{c}{\textbf{Supervised Performance}} \\
     & \multicolumn{2}{c}{\textbf{Validation}} & \multicolumn{2}{c|}{\textbf{Test}} & \multicolumn{2}{c}{\textbf{Validation}} & \multicolumn{2}{c}{\textbf{Test}} \\
    & \textbf{MRR@5} & \textbf{P@5} & \textbf{MRR@5} & \textbf{P@5} & \textbf{MRR@5} & \textbf{P@5} & \textbf{MRR@5} & \textbf{P@5} \\
    \hline
    BM25  & 0.2450 & 0.1668 & 0.2528 & 0.1642 & 0.2565 & 0.1772 & 0.2637 & 0.1755  \\
    Dense-BERT & 0.0709 & 0.0382 & 0.0726 & 0.0375 & 0.4555& 0.3155 & 0.4325 & 0.3058 \\ 
    BERT-LXMERT & 0.0744 & 0.0376 & 0.0665 & 0.0345 & 0.4704& 0.3364 & 0.4526& 0.3329 \\
    \hline
    Pre-trained BERT-LXMERT & \textbf{0.3716$^*$} & \textbf{0.2629$^*$} & \textbf{0.3364$^*$} & \textbf{0.2303$^*$} & \textbf{0.5557$^*$} & \textbf{0.4195$^*$} & \textbf{0.5603$^*$} & \textbf{0.4274$^*$} \\
    \quad \% rel. imp. w.r.t. the best baseline & \textcolor{teal}{\textbf{51.6\% $\uparrow$}} & \textcolor{teal}{\textbf{57.6\% $\uparrow$}} & \textcolor{teal}{\textbf{33.0\% $\uparrow$}} & \textcolor{teal}{\textbf{40.2\% $\uparrow$}} & \textcolor{teal}{\textbf{17.5\% $\uparrow$}} & \textcolor{teal}{\textbf{20.4\% $\uparrow$}} & \textcolor{teal}{\textbf{21.2\% $\uparrow$}} & \textcolor{teal}{\textbf{26.9\% $\uparrow$}} \\
    \hline
    \end{tabular}
    \label{tab:retrieval_results}
\end{table*}

\section{Experiments}

This section discusses the datasets, experiments, and results obtained in this paper.

\subsection{Experimental Setup}



\subsubsection*{\textbf{Dataset}}
In our experiments, we use the OK-VQA passage retrieval dataset \cite{okvqa-passage}, an extension to the OK-VQA dataset \cite{okvqa}. This dataset aims at evaluating passage retrieval tasks for outside-knowledge visual question answering tasks. This dataset contains 9009 questions for training, 2523 questions for validation, and 2523 for testing. As the retrieval collection, it uses the same Wikipedia dump that we use during pre-training (11M passages). 

\subsubsection*{\textbf{Pre-training and Fine-tuning setups}}
In order to pre-train the multi-modal dense passage retriever, we use a batch size of 32 on four RTX8000 GPUs, each with 49GB of GPU memory and a total of 256GB of RAM, which results in an effective batch size of 128. We utilize the Adam optimizer \cite{adam} with a learning rate of $10^{-5}$. A linear learning rate scheduler with 10\% of total training steps as warmup steps is used for pre-training. Additionally, gradient clipping with a clipping value of $1.0$ is used in the training procedure. The maximum length of passages and queries for each encoder is 384 and 20 tokens, respectively. We only train the model for one epoch on the pre-training data to avoid overfitting.

For fine-tuning on the OK-VQA training set, we follow the same training setup, but we use two epochs and a batch size of 4 on each GPU for a fair comparison with previous work \cite{okvqa-passage}, which results in an effective batch size of 16.

\subsubsection*{\textbf{Baselines and Terms of Comparison}}
We compare our models with the following baselines. (1) \textbf{BM25}: a baseline that only uses the question as the query and retrieves passages using BM25. (2) \textbf{Dense-BERT}: a dense retrieval baselines similar to DPR \cite{dpr} that uses questions as queries and is trained using the same training objective as ours. (3) \textbf{BERT-LXMERT}: a state-of-the-art asymmetric dense retrieval model \cite{okvqa-passage} that uses the exact same architecture as we introduced in Section \ref{sec:arch}. This baseline is basically our model but without being pre-trained using the generated data.


\subsubsection*{\textbf{Evaluation}}
Following \citet{okvqa-passage}, we use mean reciprocal rank (MRR) and precision with ranking cut-off of 5 as evaluation metrics. We use the two-tailed paired t-test with Bonferroni correction as the statistical significance test ($p < 0.05$). Since the OK-VQA dataset does not provide relevant judgment for passages, we assume a passage is identified to be positive if it contains an exact match (case insensitive) of a ground truth answer \cite{okvqa-passage}.

\subsection{Results}
This section presents our experimental results and analyzes the model performance to better demonstrate the impact of pre-training on OK-VQA performance. 



\begin{figure}
    \centering
    \includegraphics[trim="0cm 0cm 0cm 0.5cm",clip,width=.9\linewidth]{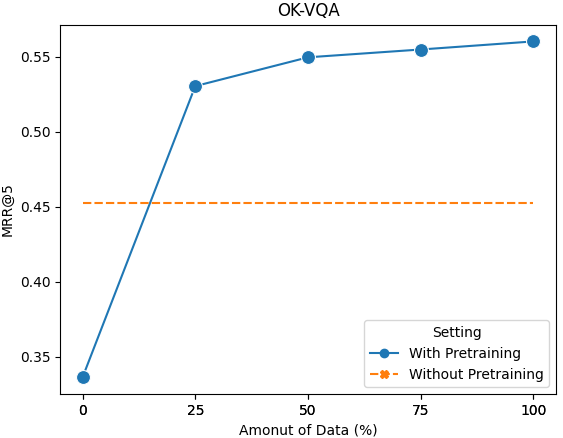}
    \caption{Learning curve for `Pre-trained BERT-LXMERT' on the OK-VQA test set. The orange line shows the performance of the BERT-LXMERT model without pre-training that is fine-tuned on 100\% of the supervised OK-VQA training data.} 
    \label{fig:learning-curve}
\end{figure}

\subsubsection*{\textbf{Zero-Shot Performance}}
In the first set of experiments, we evaluate the zero-shot capabilities of the models. In this setting, the BM25 baseline uses the default parameters ($k_1 = 1.2, b = 0.75$), and the baselines with BERT and LXMERT use the parameters learned through their (vision-) language model pre-training. The `Pre-trained BERT-LXMERT' model is trained on the data that we automatically generated. The results are reported in Table \ref{tab:retrieval_results}. BM25 demonstrates the strongest zero-shot performance. This suggests that the initialized parameters of BERT and LXMERT are not suitable for retrieval tasks. This is inline with findings by previous work on text retrieval \cite{izacard2021distilling, 10.1145/3397271.3401325}. The proposed pre-training pipeline significantly outperforms all the baselines and leads to 33\% and 40.2\% MRR@5 and P@5 improvements compared to BM25, respectively. 


\subsubsection*{\textbf{Supervised Performance}}
In the second set of experiments, we fine-tune the same models on the OK-VQA training set. All neural models use the same training procedure. The BM25 parameters are tuned through exhaustive grid search where $k_1 \in [0.5,1.5]$ and $b \in [0.2,0.8]$ with a step size of $0.2$. The model with the best MRR@5 on the validation set is selected. The selected parameters are $k_1 = 1.1, b=0.4$. In Table \ref{tab:retrieval_results}, we observe that, as expected, all neural models largely benefit from fine-tuning on the OK-VQA training set and substantially outperform BM25. Fine-tuning BERT-LXMERT that is pre-trained using the proposed data generation pipeline leads to 21.2\%  MRR@5 and 26.9\% P@5 improvements compared to BERT-LXMERT without pre-training (i.e., the current SOTA model on passage retrieval for OK-VQA \cite{okvqa-passage}).

\begin{figure}
    \centering
    \includegraphics[width=\linewidth]{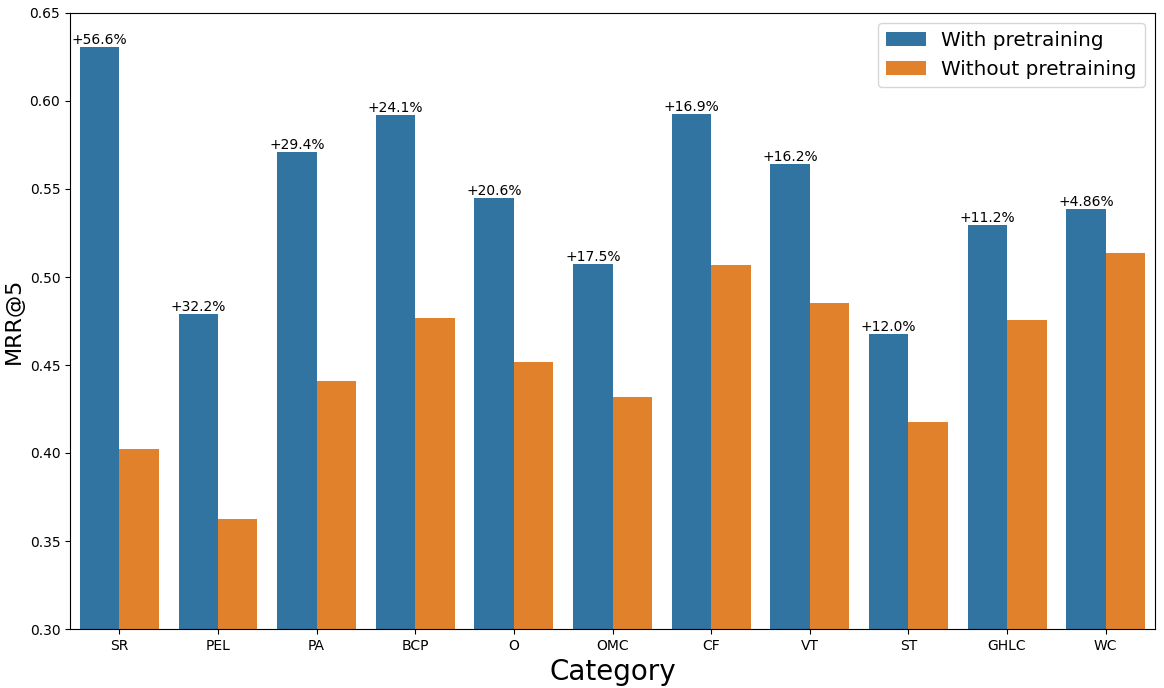}
    \caption{MRR@5 for the fine-tuned BERT-LXMERT model with and without pre-training for different question categories. The percentages on top of the bars indicate relative improvements compared to the model without pre-training.} 
    \label{fig:per-category-mrr}
\end{figure}

\subsubsection*{\textbf{Learning Curve}}
We hypothesize that the proposed pre-training pipeline reduces the need for large-scale supervised training data, which is often difficult or expensive to obtain. To validate this hypothesis, we fine-tuned our pre-trained model using 25\%, 50\%, 75\%, and 100\% of the supervised data randomly sampled from the OK-VQA training set. The results are plotted in Figure \ref{fig:learning-curve}. For the sake of space, the performance based on MRR@5 on the OK-VQA test set is reported. Other curves follow a similar behavior. The dashed orange line in the figure shows the performance of the BERT-LXMERT model without pre-training that is trained on 100\% of the OK-VQA training set. The curve demonstrates that our pre-trained model outperforms the model without pre-training by only observing 25\% of supervised training data. Moreover, the performance of the pre-trained model becomes relatively stable after observing 50\% of the supervised data, which shows that pre-training retrieval models for OK-VQA reduces the need for supervised data.

\subsubsection*{\textbf{Result Analysis.}}


To have a deeper understanding of the proposed pre-training impact on OK-VQA tasks, Figure \ref{fig:per-category-mrr} presents MRR@5 obtained by the fine-tuned BERT-LXMERT model with and without pre-training for each question category. The categories are borrowed from the OK-VQA \cite{okvqa} dataset.\footnote{The categories include ``Plants \& Animals (PA),'' ``Science \& Tech (ST),'' ``Sport \& Recreation (SR),'' ``Geography \& History \& Language \& Culture, (GHLC)'' ``Brands \& Companies \& Products (BCP),'' ``Vehicles \& Transportation (VT),'' ``Cooking \& Food (CF),'' ``Weather \& Climate (WC),'' ``People \& Everyday Life (PEL),'' ``Objects \& Material \& Clothing (OMC),'' and ``Other (O).''}  We observe that pre-training improves the OK-VQA performance on all question categories, however, the improvements are not the similar across categories. It can be seen that the highest improvement is achieved for the ``sports \& recreation'' category (56.6\%), while the lowest improvement is observed for the ``weather \& climate'' category (4.86\%). The reason is that we use the MS COCO dataset \cite{mscoco} as the image collection for automatic creation of our pret-training data and MS COCO does not include any category related to ``weather \& climate,'' ``science \& Tech,'' and ``Geography \& History \& Language \& Culture''. As a result, the extent of improvement is smaller for these categories in the OK-VQA dataset. On the other hand, a considerable proportion of images in the MS COCO dataset are related to the categories such as ``Sport \& Recreation,'' ``People \& Everyday Life,'' and ``Plants \& Animals'' that observe the highest improvements. This analysis demonstrates that including the nature of data included in the automatic data creation pipeline directly impact the downstream OK-VQA performance, and including images from underrepresented categories is likely to further improve the performance.


\section{Conclusions and Future Work}

This paper introduced a pipeline for pre-training dense retrievers for OK-VQA tasks. The proposed pipeline started from an image collection and paired each image with a passage from a knowledge source. Then, a question generation model was used to generate questions for all possible answers to the questions about the image and the passage. Finally, low-quality questions were filtered out, and negative samples for the remaining questions were selected. Our experiments suggest statistically significant improvements compared to state-of-the-art asymmetric dense retrieval performance for OK-VQA tasks.

Even though our results show consistent improvement in the OK-VQA dataset, there might be some other kinds of knowledge-intensive VQA datasets, such as FVQA \cite{fvqa}, that this pre-training approach needs to be revised. In the future, we intend to extend our data generation pipeline to other knowledge-intensive vision-language tasks. This paper also limits multi-modality to multi-modal queries and textual passages. Providing a solution for removing the mentioned limitations can be investigated in future work.

\section*{Acknowledgment}

This work was supported in part by the Center for Intelligent Information Retrieval, in part by Lowes, and in part by NSF grant \#2106282. Any opinions, findings and conclusions or recommendations expressed in this material are those of the authors and do not necessarily reflect those of the sponsor.

\bibliographystyle{ACM-Reference-Format}
\balance
\bibliography{XX-references}

\end{document}